\documentstyle[aps,psfig,floats,epsf,twocolumn,prb]{revtex}

\begin{document}
\tighten
\draft
\twocolumn[
\hsize\textwidth\columnwidth\hsize\csname @twocolumnfalse\endcsname  

\title{Nonresonant inelastic light scattering in the Hubbard model}
\author{J.K. Freericks$^*$, T.P. Devereaux$^\dagger$, R. Bulla$^\ddagger$,
and Th. Pruschke$^\ddagger$}
\address{$^*$Department of Physics, Georgetown University, Washington, DC
20057, U.S.A.}
\address{$^\dagger$Department of Physics, University of Waterloo, 
ON N2L 3G1 Canada}
\address{$^\ddagger$Theoretische Physik III, Elektronische Korrelationen und 
Magnetismus, Institut f\"ur Physik, Universit\"at Augsburg,  D-86135 Augsburg,
Germany}
\date{\today}
\maketitle

\widetext
\begin{abstract}
Inelastic light scattering from electrons
is a symmetry-selective probe of the charge
dynamics within correlated materials.  Many measurements have been made
on correlated insulators, and recent exact solutions in large dimensions
explain a number of anomalous features found in experiments.  Here we
focus on the correlated metal, as described by the Hubbard model away
from half filling.  We can determine the $B_{\rm 1g}$ Raman response
and the inelastic X-ray scattering along the Brillouin zone diagonal
exactly in the large dimensional limit.  We find a number of interesting
features in the light scattering response which should be able to be seen in 
correlated metals such as the heavy fermions.
\end{abstract}
]
\narrowtext

\section{Introduction}

Inelastic light 
scattering, and more specifically, electronic Raman and inelastic
X-ray scattering,
are useful probes of the two-particle charge
excitations within a correlated metal.  Experiments are extremely
difficult to carry out, and so far, the only correlated metals that have 
been extensively studied are the high-temperature superconductors\cite{%
irwin,hackl,uiuc}.
An experimental challenge is 
to have a large enough energy window for the inelastic light scattering
to be able to see the higher-energy charge-transfer excitations as well as
low-energy particle-hole excitations.
One class of promising materials that
could yield interesting Raman scattering results is the heavy-fermion
compounds that have sharply renormalized fermi temperatures, which compress
the charge transfer
excitations to a lower energy range (and thereby require a smaller
energy window for the scattering experiments).

On the theoretical side, much progress has been made in examining the
inelastic light scattering of correlated materials.  The first breakthrough
occurred over a decade ago when Shastry
and Shraiman (SS) \cite{ss} suggested a simplification to the theory of
nonresonant Raman scattering 
where in certain cases the nonresonant Raman response
may be equal to the optical conductivity multiplied by the frequency. Their
conjecture was recently proved\cite{freericks_devereaux_long} on a 
hypercubic lattice in $d\rightarrow\infty$
for $B_{1g}$ polarization orientations. Subsequent work on the Hubbard
model\cite{jkf_hubbard} and on the Falicov-Kimball model\cite{raman_us,%
freericks_devereaux_long}
has shown that Raman scattering is model independent in
the insulating phase, and possesses all of the anomalous features seen
in experiment on correlated insulators.

In this contribution, we focus on the correlated metal, rather than the
metal-insulator transition, and we study the behavior of the Raman response
and the inelastic X-ray scattering for a wide variety of electron
concentrations and interaction strengths.  A number of features arise
in the Raman response that are quite interesting.  Of course, the results
for the nonresonant $B_{\rm 1g}$ Raman response are similar to those
that have been shown for the optical conductivity\cite{%
jarrell_kondo,pruschke_jarrell_freericks}, but we discuss a 
number of novel features, including the inelastic scattering of X-rays
here.
In Section II we present our formalism, results appear in Section III, and
we conclude in Section IV.

\section{Formalism}

Unlike an optical conductivity experiment, which probes just one symmetry of 
the two-particle charge excitations, inelastic light
scattering can probe a number
of different symmetries by placing polarizers on the incident and
reflected light.  Three common symmetries that are examined are 
$A_{\rm 1g}$ which has the full symmetry of the lattice, $B_{\rm 1g}$
which is a d-wave symmetry, and $B_{\rm 2g}$ which is another
d-wave symmetry (rotated by $45^{\rm o}$).  The Raman response that is related
to the optical conductivity by the Shastry-Shraiman relation\cite{%
ss,freericks_devereaux_long} is the
$B_{\rm 1g}$ response. Furthermore,
the nonresonant Raman scattering in the $B_{\rm 2g}$ channel 
vanishes~\cite{freericks_devereaux_long} (for nearest
neighbor only hopping), and in the $A_{\rm 1g}$ channel
it requires knowledge of the local irreducible charge vertex (which has
never been calculated for the 
Hubbard model), so we provide results only for the $B_{\rm 1g}$ sector here.

The Hubbard model\cite{hubbard} is the simplest model of electronic correlations
that possesses a fermi-liquid fixed point for a wide range of parameter
space\cite{kotliar_mit}.  The fermi-liquid forms at low temperature, and often
one finds quite anomalous behavior in the transport
at higher temperatures, where the fermi-liquid
coherence has not yet been established\cite{jarrell_kondo}.

The Hubbard Hamiltonian~\cite{hubbard} contains two terms:
the electrons can hop between nearest neighbors
[with hopping integral $t^*/(2\sqrt{d})$ on a $d$-dimensional hypercubic
lattice~\cite{metzner_vollhardt}],
and they interact via a screened Coulomb interaction $U$ when they sit on the
same site.  All energies are measured in units of $t^*$.  The Hamiltonian is
\begin{equation}
H =-\frac{t^*}{2\sqrt{d}}\sum_{\langle i,j\rangle, \sigma}c_{i\sigma}^{\dagger}
c_{j\sigma} +
U\sum_in_{i\uparrow}n_{i\downarrow},
\label{eq: ham}
\end{equation}
where $c_{i\sigma}^{\dagger}$ $(c_{i\sigma})$ is the creation
(annihilation) operator for an electron at lattice site $i$ with spin
$\sigma$ and $n_{i\sigma}=c_{i\sigma}^{\dagger}c_{i\sigma}$ is the
electron number operator.  We adjust a chemical potential
$\mu$ to fix the average filling of the electrons ($\mu=U/2$ at half filling
of $\rho_e=1.0$ for the electrons). 

We treat the model (\ref{eq: ham}) in the infinite-dimensional limit
so the electronic properties
can be determined by dynamical mean field theory.
The starting point is
the locality of the single-particle self energy, i.e.\ the local Green function
is the Hilbert transformation of the noninteracting
density of states\cite{kotliar_mit,pruschke_jarrell_freericks}
[$\rho(\epsilon)=\exp(-\epsilon^2)/\sqrt{\pi}$ on
a hypercubic lattice in $d\rightarrow\infty$]
\begin{equation}
G(\omega)=\int_{-\infty}^{\infty}d\epsilon\rho(\epsilon)\frac{1}{
\omega+\mu-\Sigma(\omega) -\epsilon}
\label{eq: gdef}
\end{equation}
with $\omega$ approaching the real axis from above.

Formally, $G(\omega)$ can be viewed as the Green's function of an effective
single impurity Anderson model.\cite{kotliar_mit,pruschke_jarrell_freericks}
The propagator $G_0(\omega)$ for the corresponding effective non-interacting
impurity model is then determined by
\begin{equation}
G_0^{-1}(\omega)=G^{-1}(\omega)+\Sigma(\omega).
\label{eq: g0def}
\end{equation}
Next an Anderson impurity model solver must be employed to determine
the local Green's function from $G_0(\omega)$ and $U$.
Here, we use the numerical renormalization group (NRG) technique\cite{alex,NRG}
to calculate the
self energy on the real axis, and then employ Eq.~(\ref{eq: gdef})
to determine the new Green's function.  This algorithm is iterated until the
Green's functions converge to a fixed point.

The NRG is based on a logarithmic discretization of the
energy axis, i.e.\ one introduces a parameter $\Lambda>1$ and divides
the energy axis into intervals $\pm[\Lambda^{-(n+1)},\Lambda^{-n}]$ for
$n=0,1,\ldots,\infty$ \cite{alex,NRG}. With some further manipulations
 one can map the original model onto a semi-infinite chain, which
can be solved  iteratively by starting from the impurity and successively
adding chain sites. Since the coupling between two adjacent sites $n$ and
$n+1$ vanishes  like $\Lambda^{-n/2}$ for large $n$, the low-energy
states of the chain with $n+1$ sites are determined by a comparatively small
number $N_{\rm states}$ of states close to the ground state of the $n$-site
system. In practice, one retains only these $N_{\rm states}$ from the $n$-site
chain to set up the Hilbert space for $n+1$ sites and thus prevents the usual
exponential growth of the Hilbert space as $n$ increases. Eventually, after
$n_{\rm NRG}$ sites have been included in the calculation, adding another site
will not change the spectrum significantly and one terminates the calculation.

It is obvious, that for any $\Lambda>1$ the NRG constitutes an approximation
to the system with a continuum of band states but becomes exact in the
limit $\Lambda\to 1$. Performing this limit is, of course, not possible as one
has to simultaneously increase the number of retained states to infinity. One
can, however, study the  $\Lambda$- and $N_{\rm states}$-dependence of the NRG
results and perform the  limit $\Lambda\to 1$, $N_{\rm states}\to\infty$ by
extrapolating these data. Surprisingly one finds that the dependence of the
NRG results on $\Lambda$ as well as on the cut-off $N_{\rm states}$ is
extremely mild; in  most cases, a choice of $\Lambda=2$ and
$N_{\rm states}=300\ldots500$ is sufficient.

While the knowledge of the states is sufficient to calculate
thermodynamic properties, the discretization leads to a Green's
function consisting of a discrete set of poles and an appropriate
coarse-graining  procedure has to be applied.\cite{Sak89,bulla}.
Obviously, the NRG is most accurate for energies near the chemical
potential, where the logarithmic grid is finest.
It
is less accurate
at energies far away from the chemical potential,
because the energy grid is coarser and one uses an asymmetric (logarithmic)
broadening for the delta functions.
Since our final response functions involve complex
convolution integrals containing the
interacting Green function,
the results are expected to be most accurate in the low-frequency region.
At high frequencies one may still expect the qualitative features to be
correct, but details like the precise distribution of spectral weight will
typically be less reliable. The quantitative errors in both the Green's function
and the inelastic light scattering response functions, however, are difficult 
to estimate.

The nonresonant Raman response in the B$_{\rm 1g}$ channel 
and the inelastic X-ray scattering along the Brillouin-zone diagonal
(in the $B_{\rm 1g}$ channel) have
no vertex corrections~\cite{khurana,raman_us,freericks_devereaux_long} 
and are equal to the bare bubbles.  The formula for the Raman response
has been presented
elsewhere~\cite{freericks_devereaux_long} and can also be found from the
SS relation~\cite{ss,jarrell_kondo,pruschke_jarrell_freericks}---the 
imaginary part of 
the nonresonant $B_{\rm 1g}$ Raman response is  
\begin{eqnarray}
{\rm Im}\chi_{B_{\rm 1g}}({\bf q}=0,\nu)
&=&c\int d\omega [f(\omega)-f(\omega+\nu)]\cr
&\times&\int d\epsilon \rho(\epsilon) A(\epsilon,\omega)A(\epsilon,\omega+\nu),
\label{eq: raman}
\end{eqnarray}
where $f(\omega)=1/[1+\exp(\omega/T)]$ is the Fermi function, 
$A(\epsilon,\omega)=-{\rm Im}[1\{\omega+\mu-\Sigma(\omega)-\epsilon\}\pi ]$
is the spectral function, and $c$ is a constant.  

The inelastic light scattering by a photon with momentum ${\bf q}=(q,q,...,q)$
can also be calculated with the bare bubble, because one can show that all
vertex corrections vanish here as well\cite{footnote}.  The result depends
only on the parameter $X=\cos q$, and is
\begin{eqnarray}
\chi_{B_{\rm 1g}}(\textbf{q},\nu)&=&\frac{i}{4\pi}\int_{-\infty}^{\infty}
d\omega\bigl \{ f(\omega)\chi_0(\omega;X,\nu)\cr
&-&f(\omega+\nu)\chi_0^*(\omega;X,\nu)-
[f(\omega)-f(\omega+\nu)]\cr
&\times&\tilde\chi_0(\omega;X,\nu) \bigr \}\nonumber\\
\label{eq: b1g_final}
\end{eqnarray}
with
\begin{eqnarray}
\chi_0(\omega;X,\nu)&=&-\int_{-\infty}^{\infty}d\epsilon\rho(\epsilon)
\frac{1}{\omega+\mu-\Sigma(\omega)-\epsilon}\frac{1}{\sqrt{1-X^2}}\cr
&\times&F_\infty \left (
\frac{\omega+\nu+\mu-\Sigma(\omega+\nu)-X\epsilon}{\sqrt{1-X^2}}
\right ) , \label{eq: chi0}
\end{eqnarray}
and
\begin{eqnarray}
\tilde\chi_0(\omega;X,\nu)&=&-\int_{-\infty}^{\infty}d\epsilon\rho(\epsilon)
\frac{1}{\omega+\mu-\Sigma^*(\omega)-\epsilon}\frac{1}{\sqrt{1-X^2}}\cr
&\times&F_\infty \left (
\frac{\omega+\nu+\mu-\Sigma(\omega+\nu)-X\epsilon}{\sqrt{1-X^2}}
\right ) . \label{eq: chi0tilde}
\end{eqnarray}
Here we have $F_\infty(z)= \int
d\epsilon \rho(\epsilon)/(z-\epsilon)$, which is the Hilbert transform of
the noninteracting density of states.  Note that the $X=1$ limit of
Eq.~(\ref{eq: b1g_final}) corresponds to the optical photon (${\bf q}=0$)
and it reduces to the Raman scattering case in Eq.~(\ref{eq: raman}) as it
must.

\section{Results}

\begin{figure}[htbf]
\epsfxsize=2.9in
\epsffile{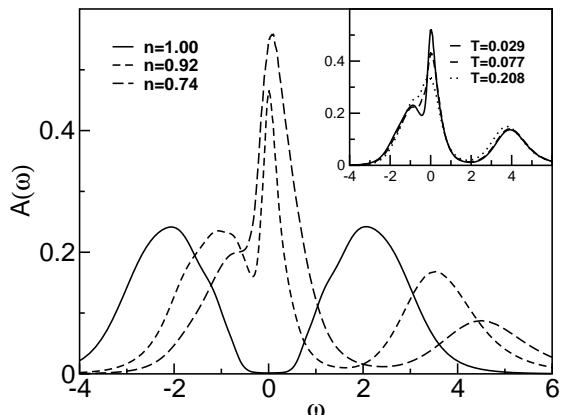}
\caption{
\label{fig: DOS}
Evolution of the single-particle DOS for $U=4.24$ with filling for $T=0.029$
(main panel) respectively with temperature for $\rho_e=0.92$ (inset). Note
especially the strong variation of the quasi-particle peak in both cases.
In addition, with decreasing filling there is a strong redistribution of
spectral weight from the upper Hubbard band to the Fermi energy. For even
lower filling, the upper Hubbard band vanishes completely.  
}
\end{figure}

\begin{figure}[htbf]
\epsfxsize=2.9in
\epsffile{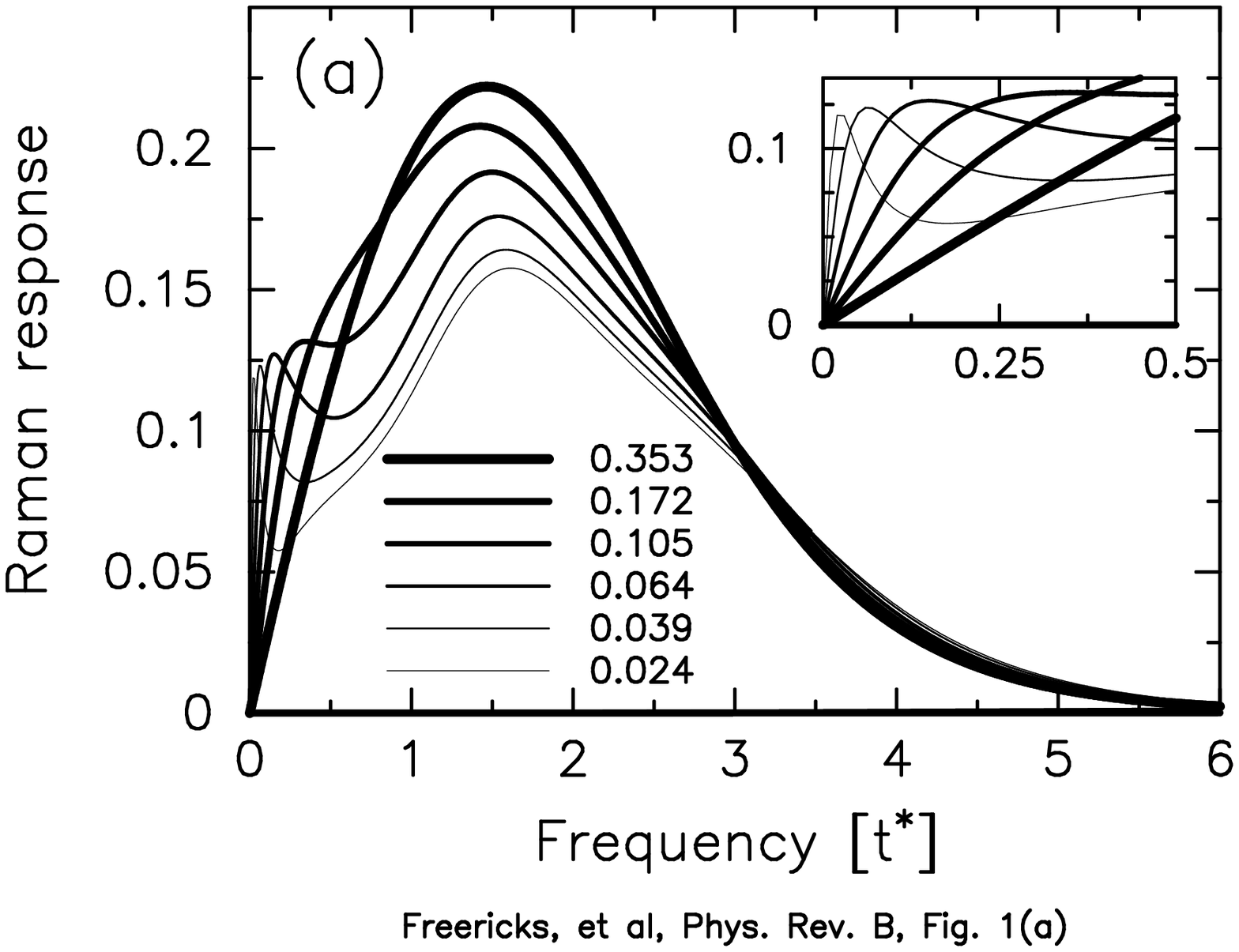}
\epsfxsize=2.9in
\epsffile{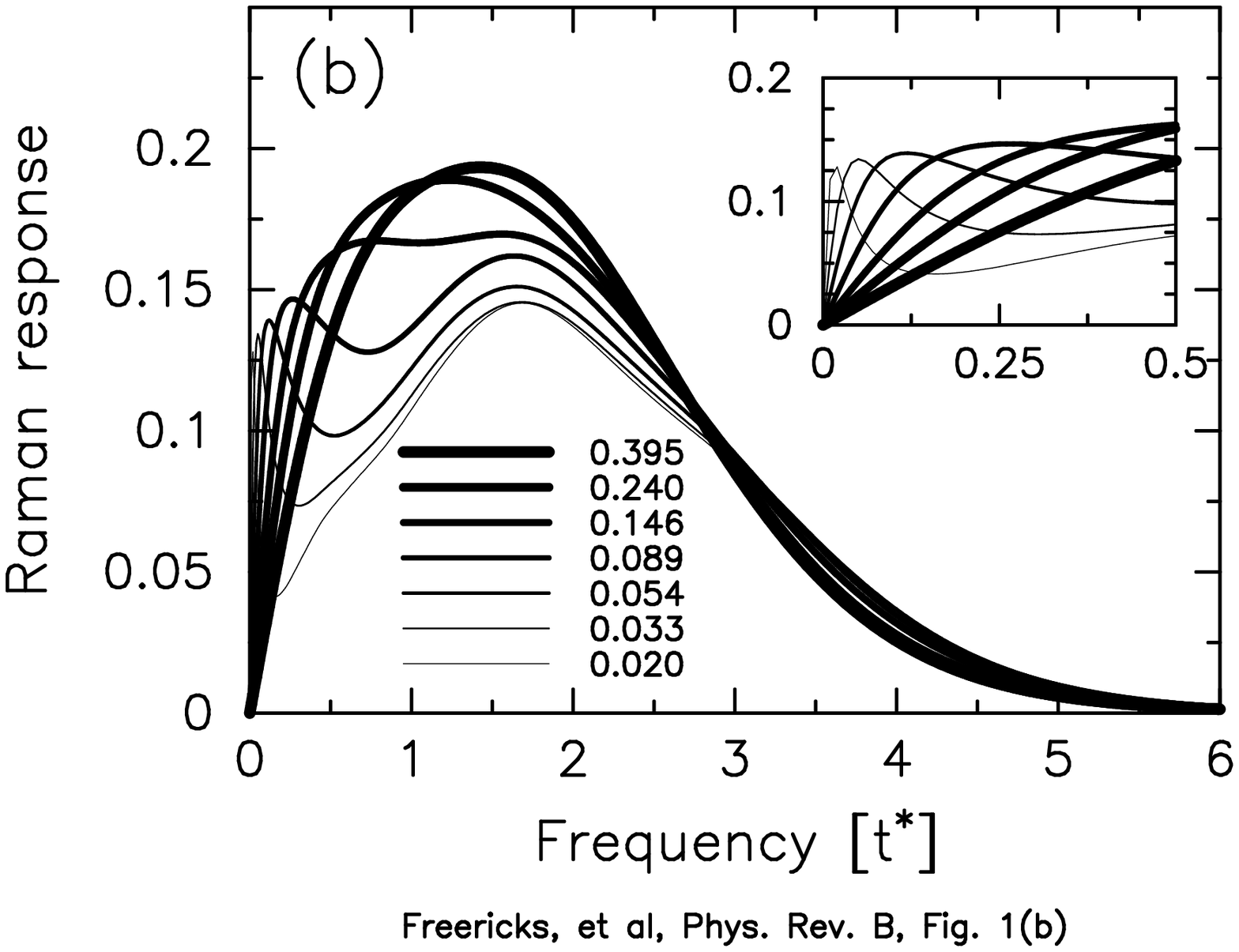}
\epsfxsize=2.9in
\epsffile{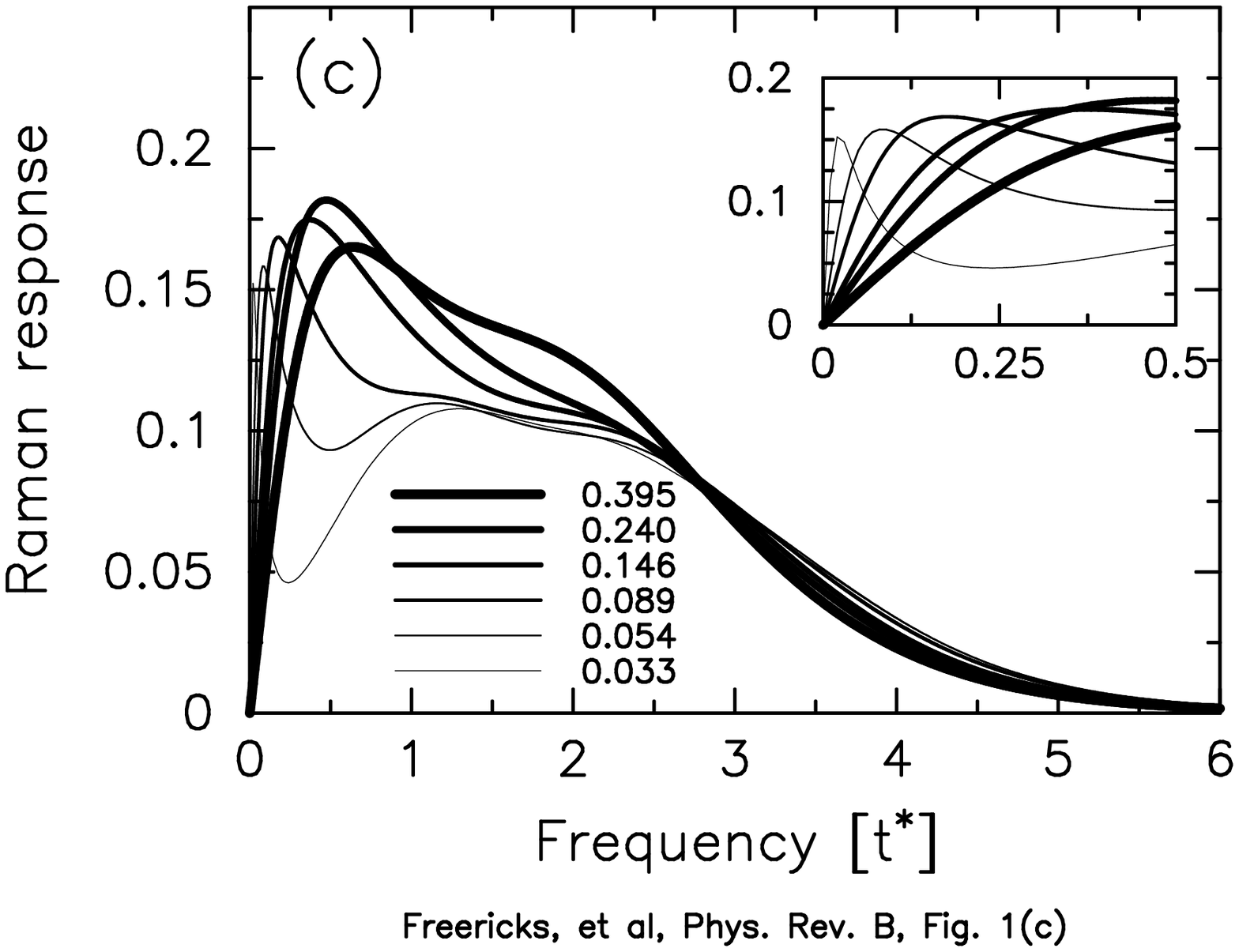}
\epsfxsize=2.9in
\epsffile{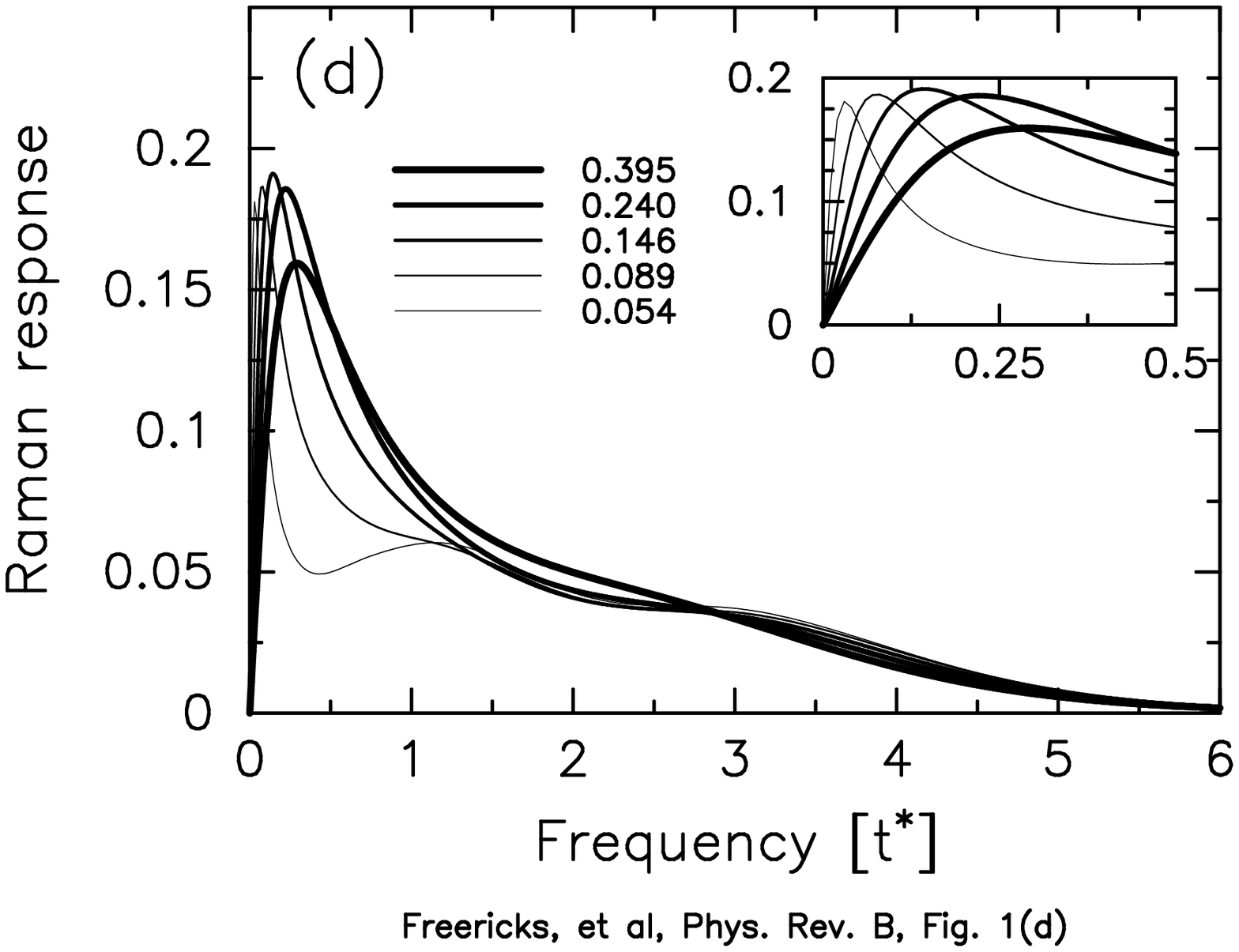}
\caption{
\label{fig: metal_n}
Raman response for different temperatures at $U=2.12$.  The different panels
refer to different electron fillings: (a) $\rho_e=1.0$;
(b) $\rho_e=0.9$; (c) $\rho_e=0.8$; and (d) $\rho_e=0.6$. 
The numbers in the legends label the temperature. Inset into each panel
is a blow-up of the low-frequency region to show the fermi-liquid peak
formation and evolution with $T$.}
\end{figure}

Before discussing the Raman spectra in detail, let us briefly look at the
single-particle density of states (DOS). Although this quantity has been
extensively discussed in the literature\cite{pruschke_jarrell_freericks}
it is instructive to review the major properties here because they will
be needed to interpret the Raman spectra.

The evolution of the DOS for
$U=4.24$ upon varying filling and temperature is
collected in Fig.~\ref{fig: DOS}. Quite generally, one can identify three
structures: A comparatively sharp peak at the Fermi energy $\omega=0$,
and two broad features, the so-called Hubbard bands, representing incoherent
(local) charge excitations (the sharp peak vanishes at $n=1.0$ because the
system is in the 
Mott-insulating state). With decreasing filling (Fig.~\ref{fig: DOS},
main panel) these latter structures become less pronounced, the lower one
being absorbed into the quasi-particle peak, and the spectral weight of the
upper one redistributed to the Fermi level. For still lower filling one would
eventually arrive at a single peak at the Fermi level, as in the non-interacting
system.
For a fixed filling, on the other hand, the variation with temperature
(Fig.~\ref{fig: DOS} inset) does not affect the incoherent structures much,
but rather leads to a strong variation in the quasi-particle
peak at the Fermi energy. Changing the value of $U$ does not lead to
qualitatively different features, only the specific values of the filling or
temperature, where changes become pronounced, will vary.

We now turn to the discussion of our results for Raman scattering by examining a
correlated metal with $U=2.12$.  This system is metallic (in the paramagnetic 
phase) for all electron fillings.  The fermi temperature is quite low though,
so the quasi-particle peak only appears in the interacting DOS for low enough
temperature (c.f.\ Fig.~\ref{fig: DOS}).
This has implications for the Raman response as well.  We show the Raman
scattering at $U=2.12$ for a variety of temperatures in Fig.~\ref{fig: metal_n}.
We plot results for four different electron concentrations (a) $\rho_e=1.0$,
(b) $\rho_e=0.9$, (c) $\rho_e=0.8$, and (d) $\rho_e=0.6$.  Inset in each panel
is a blow up of the low-temperature region where the fermi peak appears.
The behavior appears qualitively different near half filling and far from
half filling.  Near half filling, we see a large charge-transfer peak at
$\nu\approx 1.5$ present at high temperature, which has a fermi-liquid
peak separate out of it at low temperature.  The fermi peak narrows and is
pushed to lower energy as $T$ is lowered, as we expect for a correlated
metal as the phase space for particle-hole scattering is reduced.
The relative weight of the low energy (fermi-liquid) feature and the
charge-transfer feature changes as we move away from half filling.  At
$\rho_e=0.8$, we find a broad high-energy feature, which may be separating
into two peaks centered at approximately 1.5 and 2.5 at the lowest
temperature.  The scattering for $\rho_e=0.6$ is even more dramatically
modified.  We see no large charge transfer peak at high temperature,
just a broad low-energy peak, reminiscent of the fermi-liquid peak.  As
$T$ is lowered, this broad peak narrows and evolves to low energy.  As it
does so, we see two additional features emerge in the Raman response,
corresponding to a mid-infrared bump and a charge transfer bump.  Since the
interacting DOS typically has a three peak structure
(see Fig.~\ref{fig: DOS}), with a lower Hubbard
band, an upper Hubbard band, and a quasiparticle peak, these three features
arise from transitions within the quasiparticle peak, between the 
quasiparticle peak and the
lower band, and between the quasiparticle peak and the upper band.  When the 
electron filling is less than half filled, the quasiparticle
peak in the DOS lies well within
what would be termed the lower Hubbard band.  Finally, we note that the
evolution of Raman scattering with doping is an evolution toward weaker
correlations as $\rho_e$ is reduced, which is precisely what we
expect for the Hubbard model, which is most strongly correlated at
half filling.

\begin{figure}[htbf]
\epsfxsize=3.0in
\epsffile{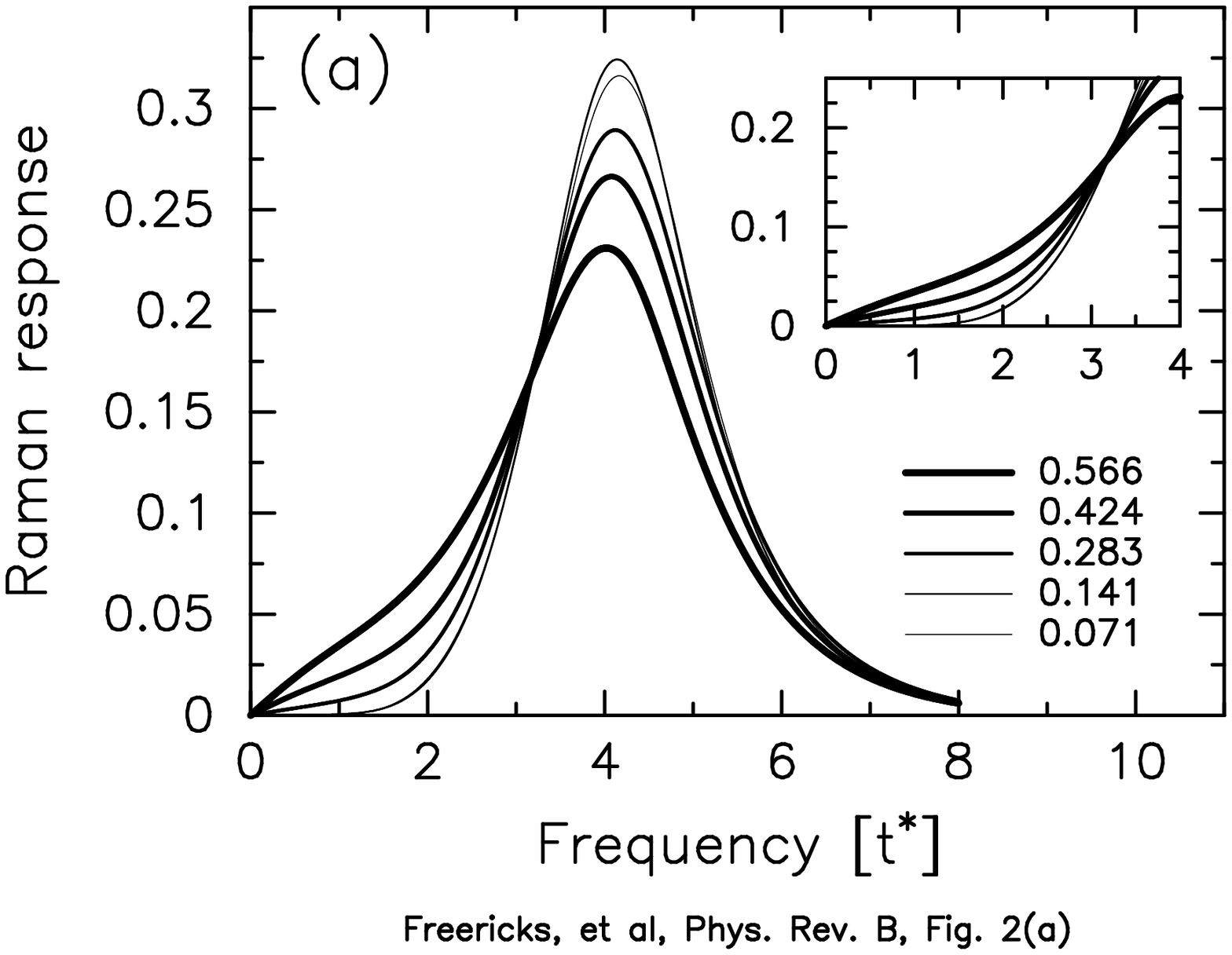}
\epsfxsize=3.0in
\epsffile{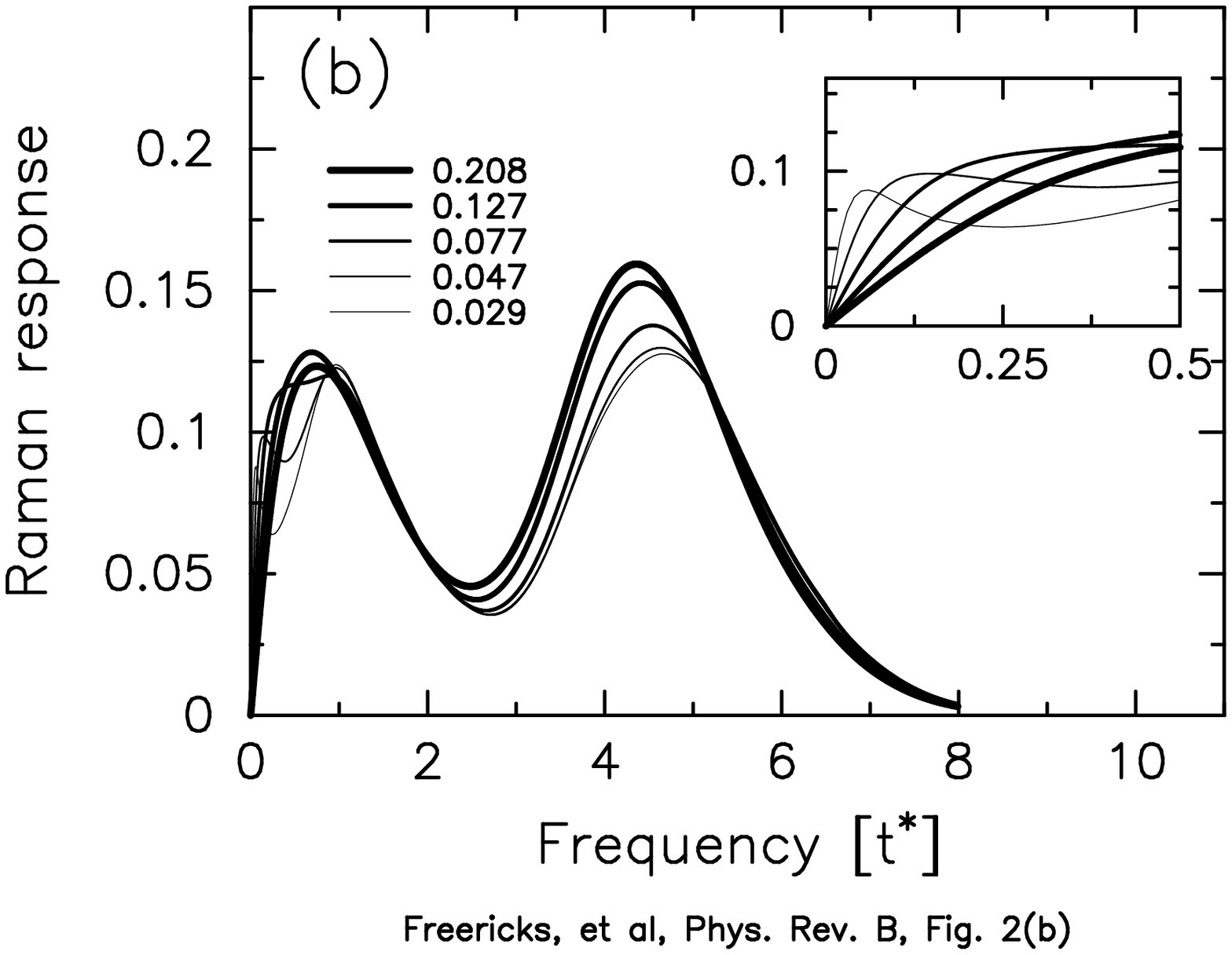}
\epsfxsize=3.0in
\epsffile{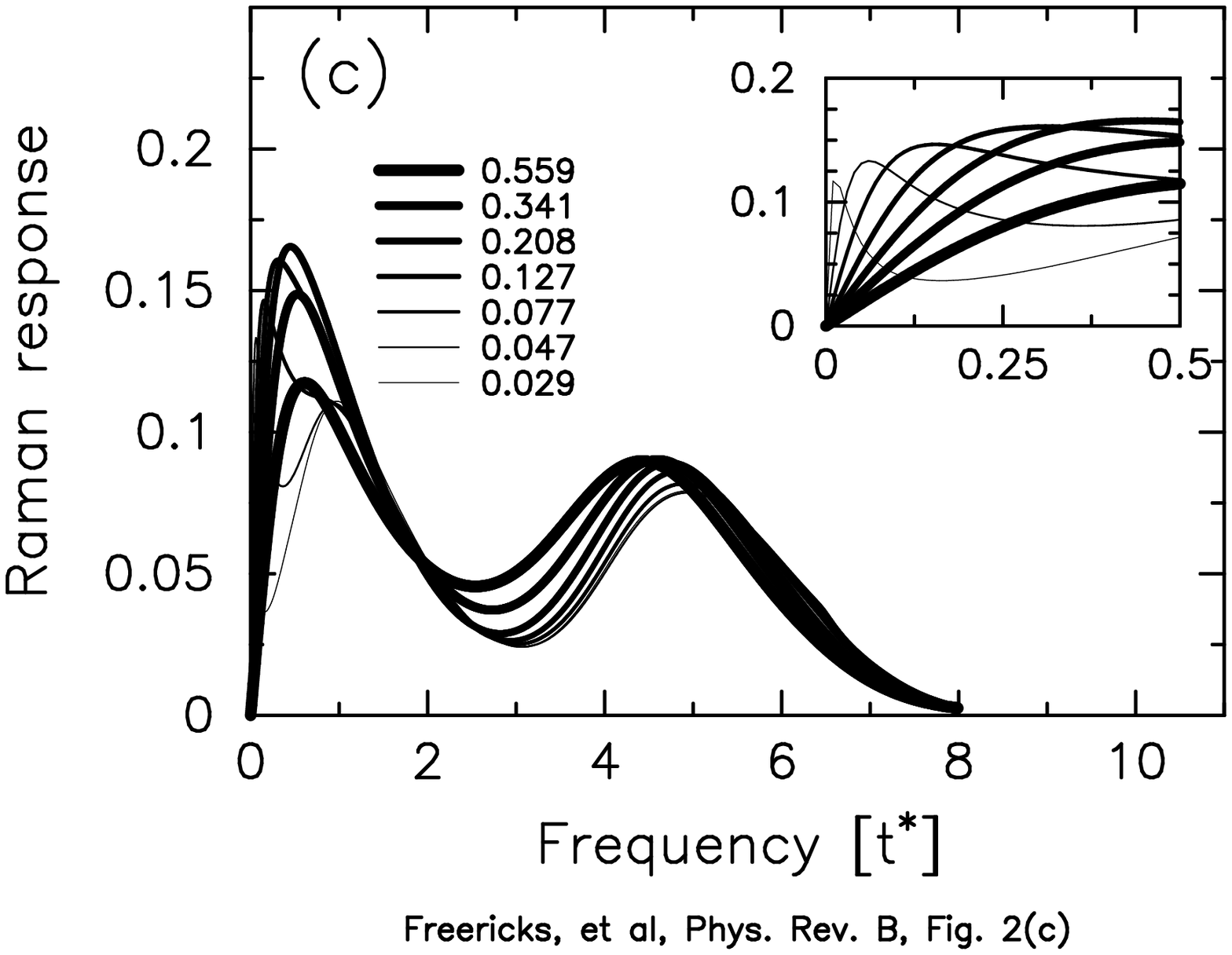}
\caption{
\label{fig: corr_n}
Raman response for different temperatures at $U=4.24$.  The different panels
refer to different electron fillings: (a) $\rho_e=1.0$;
(b) $\rho_e=0.9$; and (c) $\rho_e=0.8$. 
The numbers in the legends label the temperature. Inset into each panel
is a blow-up of the low-frequency region to show the fermi-liquid peak
formation and evolution with $T$.}
\end{figure} 

We compare these moderate correlation results to a more strongly correlated
system $(U=4.24)$
in Fig.~\ref{fig: corr_n}.  Here three different electron fillings
are shown: (a) $\rho_e=1.0$; (b) $\rho_e=0.9$, and (c) $\rho_e=0.8$.  At
half filling [panel (a)], the system is a correlated insulator, 
and the Raman response
displays the characteristic features of an insulator, including the appearance
of low-energy spectral weight at an onset temperature that is smaller than
the $T=0$ gap, and an isosbestic point, where the Raman response is independent
of temperature (here occurring near $\nu\approx 3$).  As the system is doped,
its behavior changes dramatically, because the doped system is metallic.
Here  [panels (b) and (c)] we see two peaks at high temperature.  As $T$ is
lowered, the low-energy peak has a fermi-liquid peak split off and narrow
and move to lower energy as $T$ is reduced.  The relative weight of the 
low-energy features (the fermi-liquid peak and the ``mid-IR'' peak) to
the charge-transfer peak increases as the filling moves away from half filling.
We also can see that the fermi-liquid temperature increases as we move away
from half filling because of the sharper fermi peak at the same temperature
in panel (c) versus panel (b).  This is to be expected, because as the
relative correlations are reduced, the coherence temperature should
increase.

One might be tempted to perform an analysis similar to what was done for the
optical conductivity\cite{jarrell_kondo,pruschke_jarrell_freericks}, where one 
examines
the relative weights of the fermi peak, the mid-IR bump and the charge
transfer peak.  Such an analysis is really only feasible for the situation
where one has a sum rule satisfied (like the optical conductivity).  The
multiplication of the optical conductivity by the frequency to produce the Raman
response obviously changes the sum rule, and makes such simple identifications
more difficult\cite{sum}.  
But the general qualitative feature of more scattering going
into the charge transfer peak as the correlations increase certainly still
holds.  The fermi peak is smaller in $B_{\rm 1g}$ Raman scattering, since it
is concentrated at small frequencies, and it is reduced by the extra power
of frequency that multiplies the optical conductivity to produce the Raman
result.

\begin{figure}[htbf]
\epsfxsize=3.0in
\epsffile{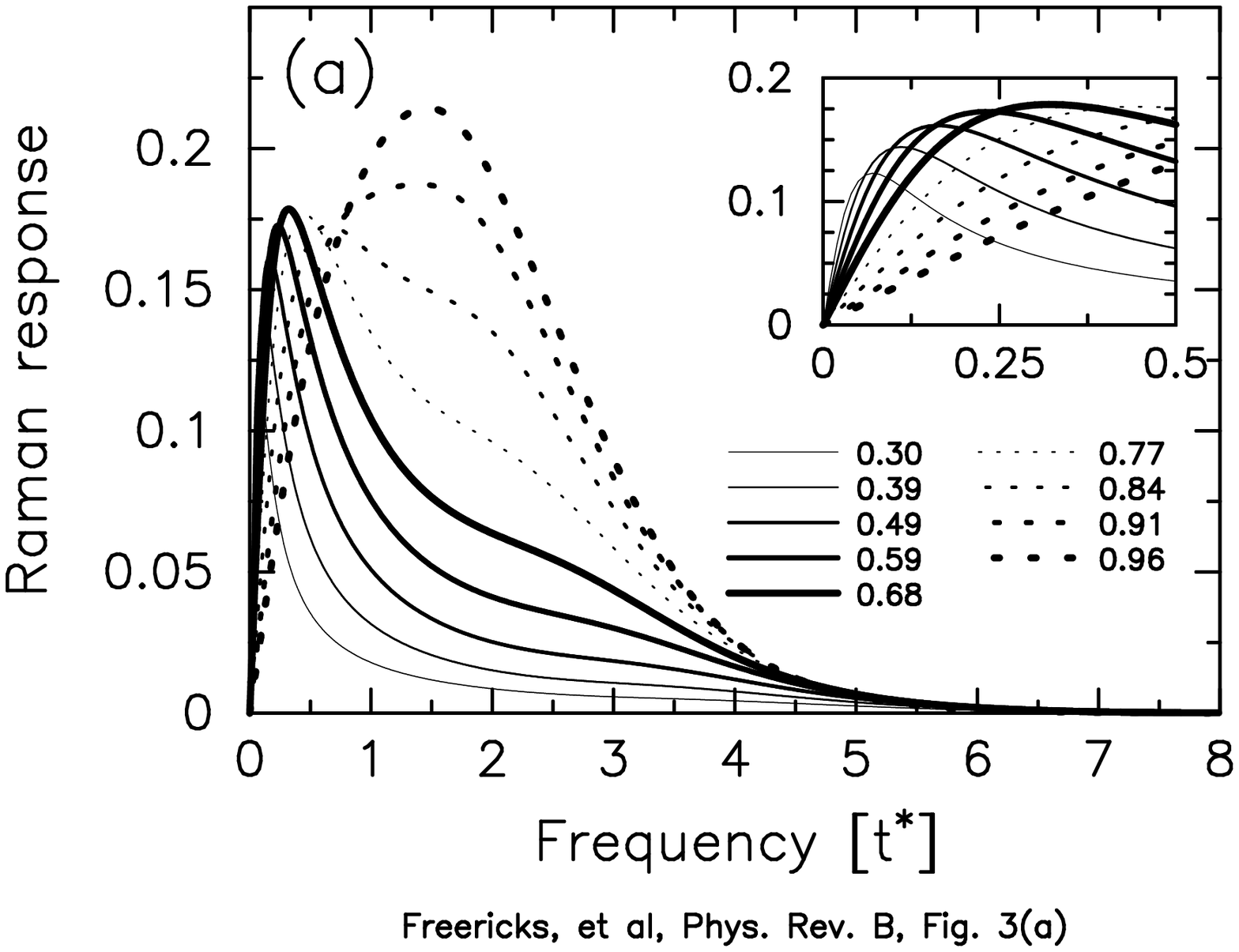}
\epsfxsize=3.0in
\epsffile{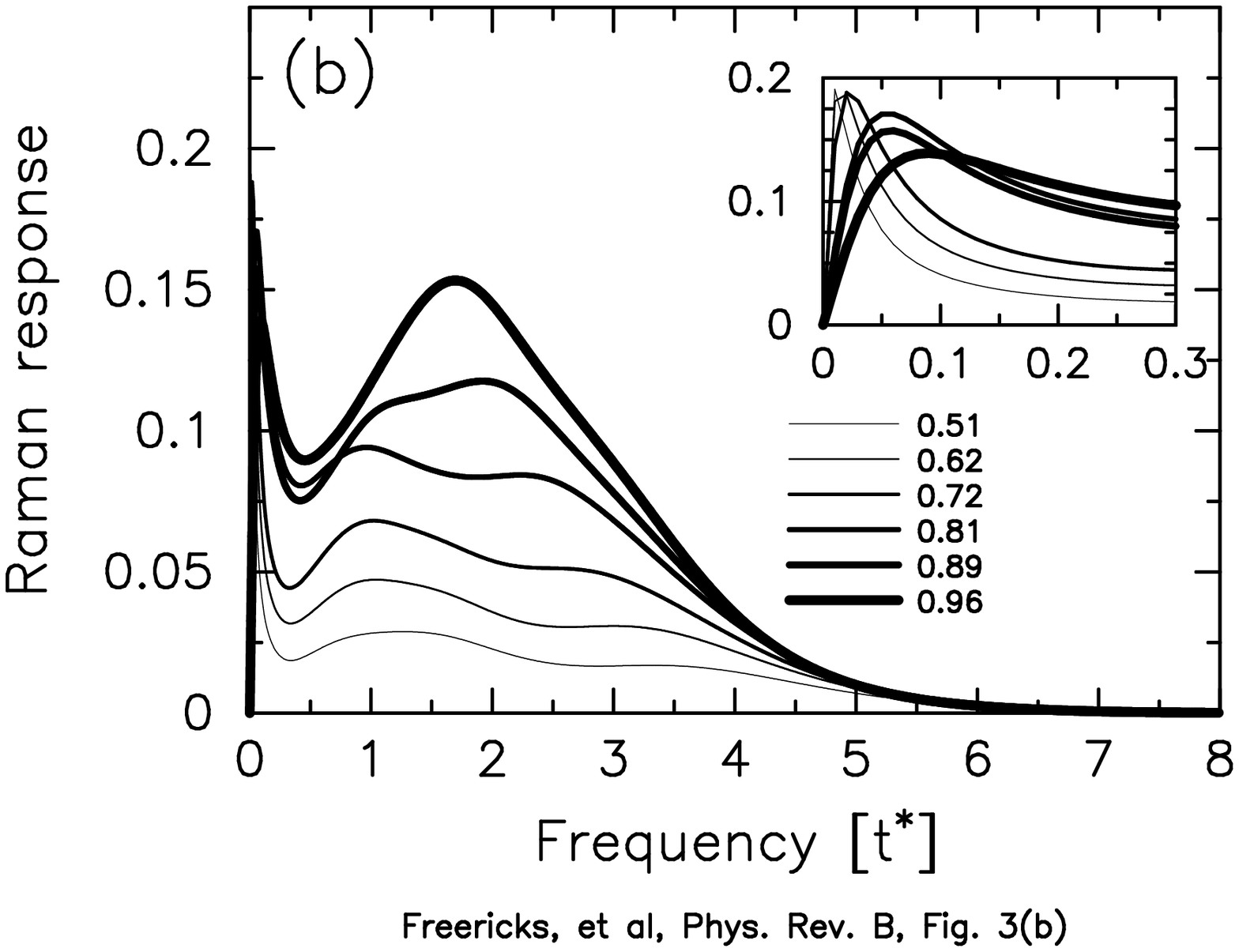}
\caption{                                                                       
\label{fig: metal_t}
Raman response for different fillings at $U=2.12$ and constant temperature.  
The different panels
refer to different $T$,  (a) $T=0.341$ and (b) $T=0.0472$.
The numbers in the legends label the electron filling.
Inset into each panel
is a blow-up of the low-frequency region to show the fermi-liquid peak
formation and evolution with electron filling.}
\end{figure}

We examine the Raman scattering at constant temperature in a moderately
correlated metal $U=2.12$
in Fig.~\ref{fig: metal_t}.  At high temperature [panel (a)],
we see a broad charge-transfer-like feature near half filling, but as the
electron concentration moves away from 1.0, we see spectral weight shift to
low energy and evolve into a sharp fermi liquid peak.  For intermediate
fillings, we can still see a remnant of the charge-transfer peak at high
energy, as a wiggle in the Raman response, but the size of that feature
decreases sharply with filling. This picture clearly shows how the fermi
temperature depends strongly on electron concentration, as the fermi temperature
evolves to become larger than $0.341$ around $\rho_e\approx 0.6$.  At low
temperature [panel (b)], 
we see a strong fermi-liquid peak, that narrows and moves to
lower energy with electron filling, as expected.  In addition, we
see the higher-energy processes are sharpened, with a ``mid-IR'' feature
occurring near $\nu\approx 1$ and a charge-transfer feature at higher
energies.  Once again, the relative weight of the higher-energy feature
gets smaller as we move the filling farther from half filling.

\begin{figure}[htbf]
\epsfxsize=3.0in
\epsffile{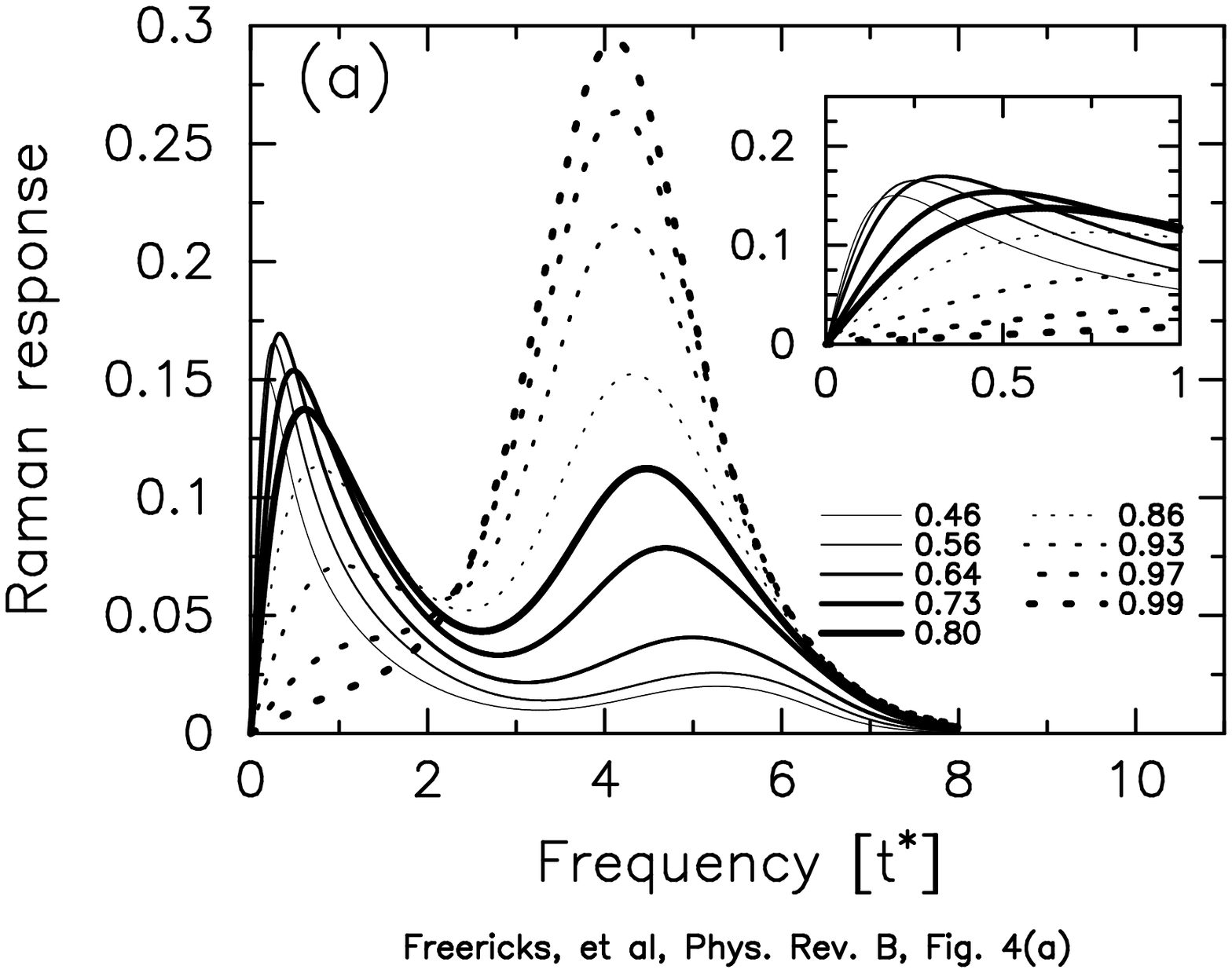}
\epsfxsize=3.0in
\epsffile{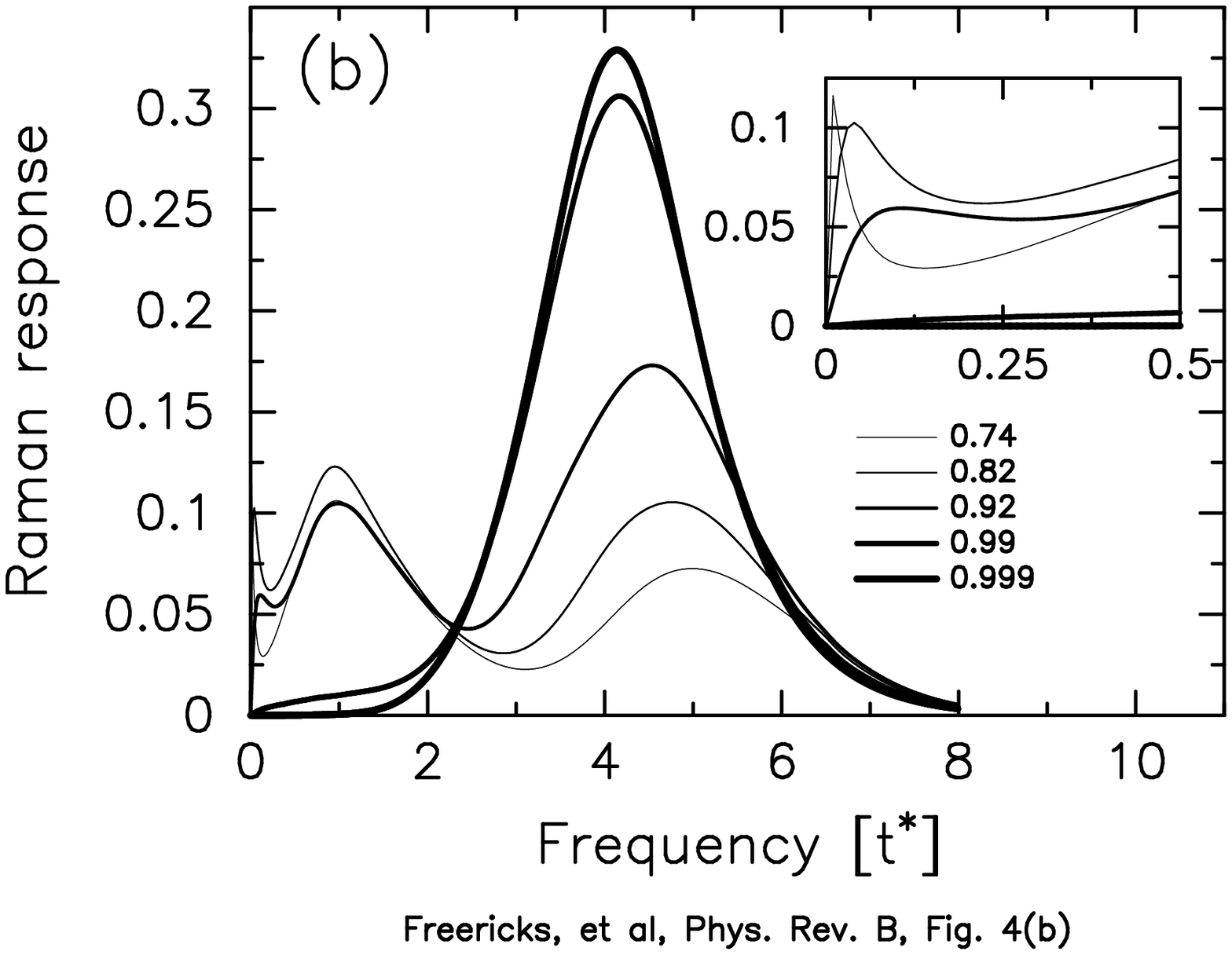}
\caption{
\label{fig: corr_t}
Raman response for different fillings at $U=4.24$ and constant temperature.
The different panels
refer to different $T$,  (a) $T=0.341$ and (b) $T=0.0248$.
The numbers in the legends label the electron filling.
Inset into each panel
is a blow-up of the low-frequency region to show the fermi-liquid peak
formation and evolution with electron filling.}
\end{figure}

In Fig.~\ref{fig: corr_t}, we plot the Raman response at fixed $T$ for
a strongly correlated system
$U=4.24$.  In the high-temperature case [panel (a)], we find two prominent 
peaks, one a charge transfer peak at high energy and one a ``mid-IR'' peak
at lower energy.  The temperature is too high to see the fermi-liquid peak
forming.  We can see, however, the dramatic transfer of relative spectral
weight to lower energy as the filling moves farther from half filling,
and the charge-transfer peak shrinks.  At the lower temperature [panel (b)],
we see a sharp charge-transfer peak, and an approximate isosbestic point, where
the Raman response is almost independent of electron filling near $\nu\approx
U/2$.  Close to the insulating phase, there is very small low-energy spectral
weight.  As we move away from half filling, the low-energy spectral weight 
rises, developing into a ``mid-IR'' feature and having an even lower-energy
fermi peak.  This result is similar to what was seen in the optical
conductivity\cite{jarrell_kondo,pruschke_jarrell_freericks}, where there 
appeared to be an isosbestic point at about $U/2$
as a function of doping for fillings close to half filling.  The isosbestic
point is approximately preserved here, because multiplying by the frequency 
does not remove such behavior.

\begin{figure}[htbf]
\epsfxsize=3.0in
\epsffile{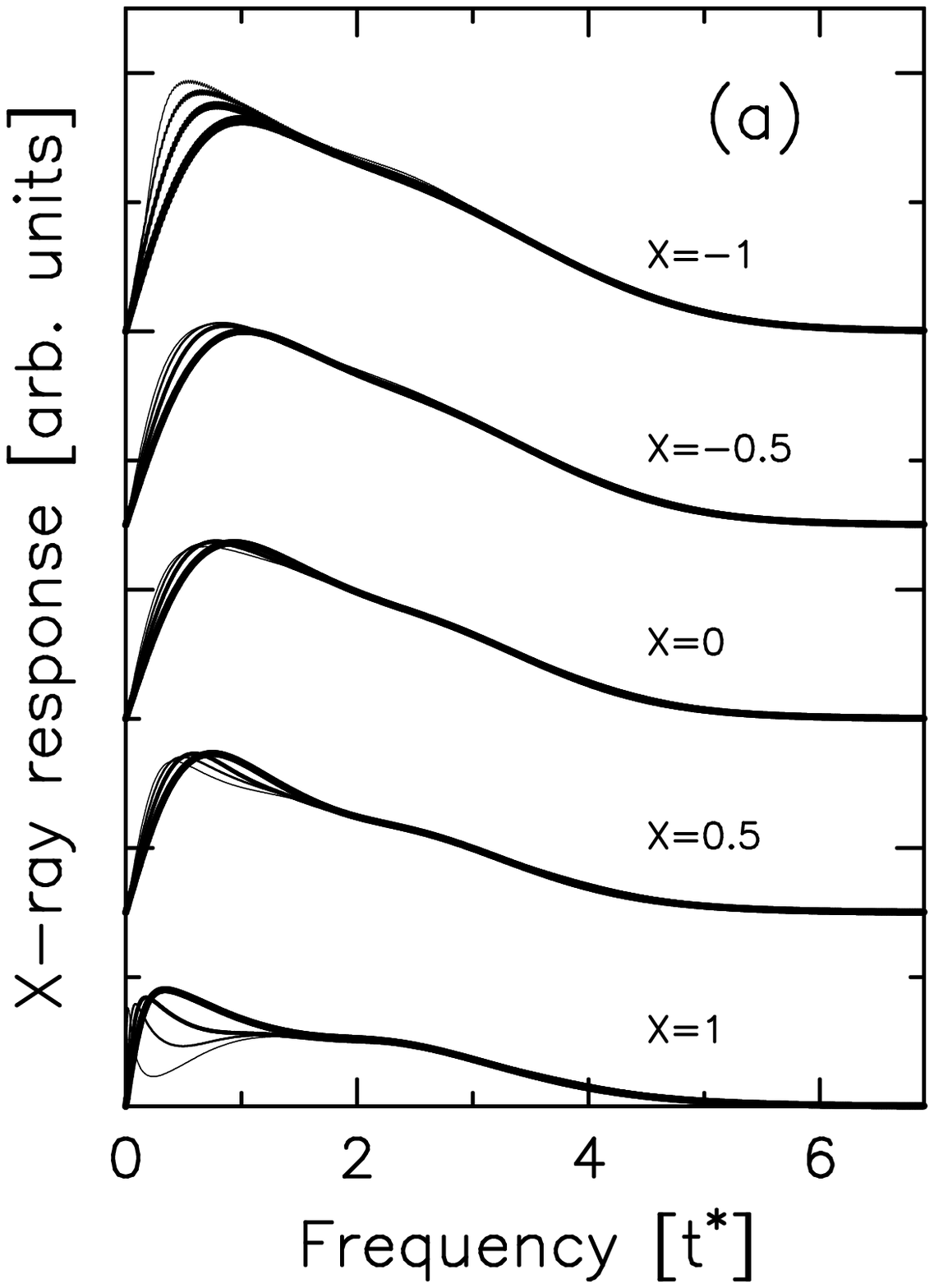}
\epsfxsize=3.0in
\epsffile{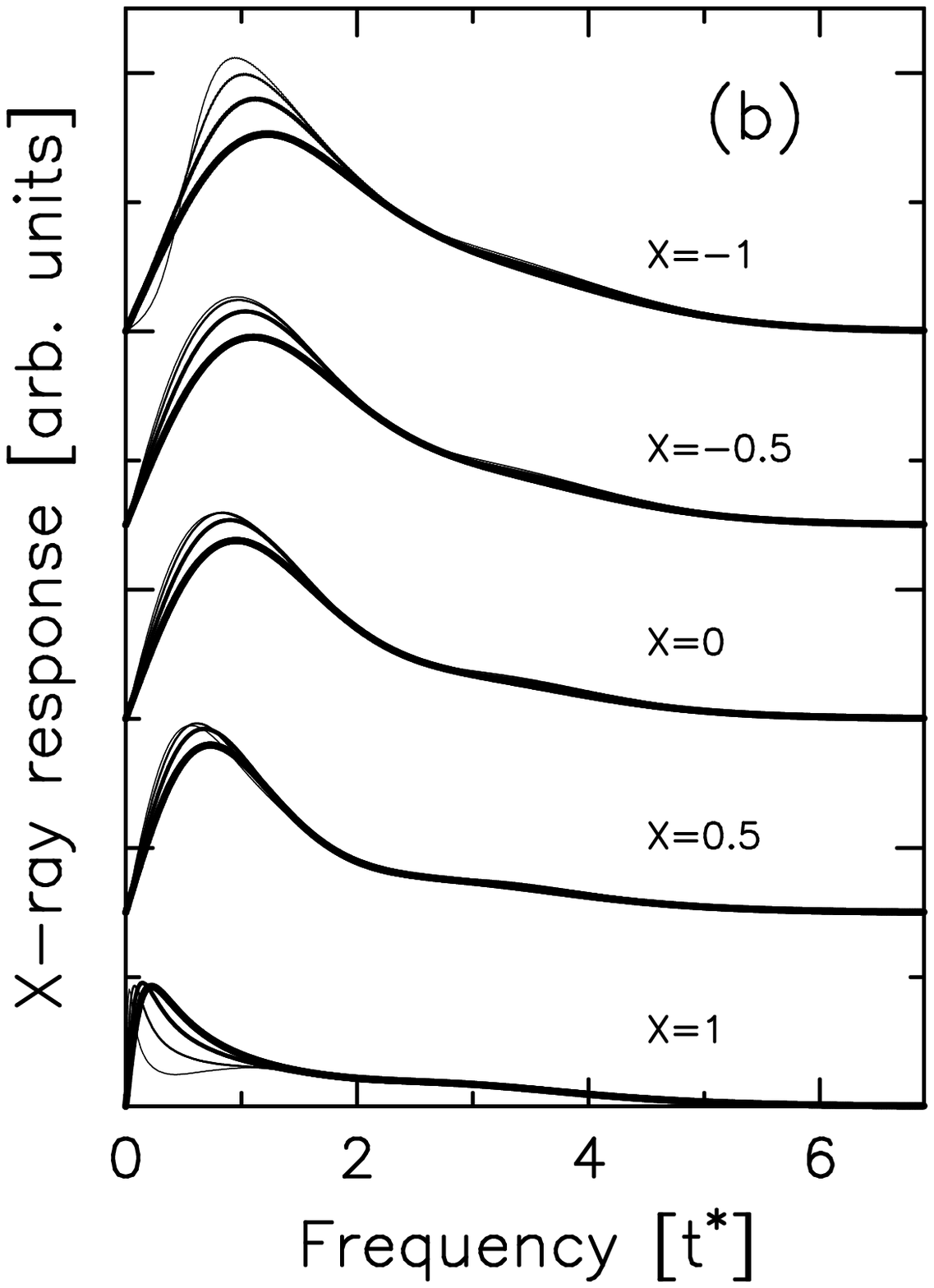}
\caption{
\label{fig: metal_xray}
Inelastic light scattering at $U=2.12$ and (a) $\rho_e=0.8$ and (b) $\rho_e=0.6$
for different temperatures.  Five values of the transferred photon momentum
are plotted, each shifted by an appropriate amount, and running from the 
zone center to the zone boundary along the zone diagonal.  The results
are for the $B_{\rm 1g}$ sector only. In panel (a) the temperature decreases
as the thickness decreases and ranges from 0.146 to 0.089 to 0.054 to 0.033.
In panel (b) the temperature also decreases as the thickness decreases
and ranges from 0.240 to  0.146 to 0.089 to 0.054.
}
\end{figure} 

\begin{figure}[htbf]
\epsfxsize=3.0in
\epsffile{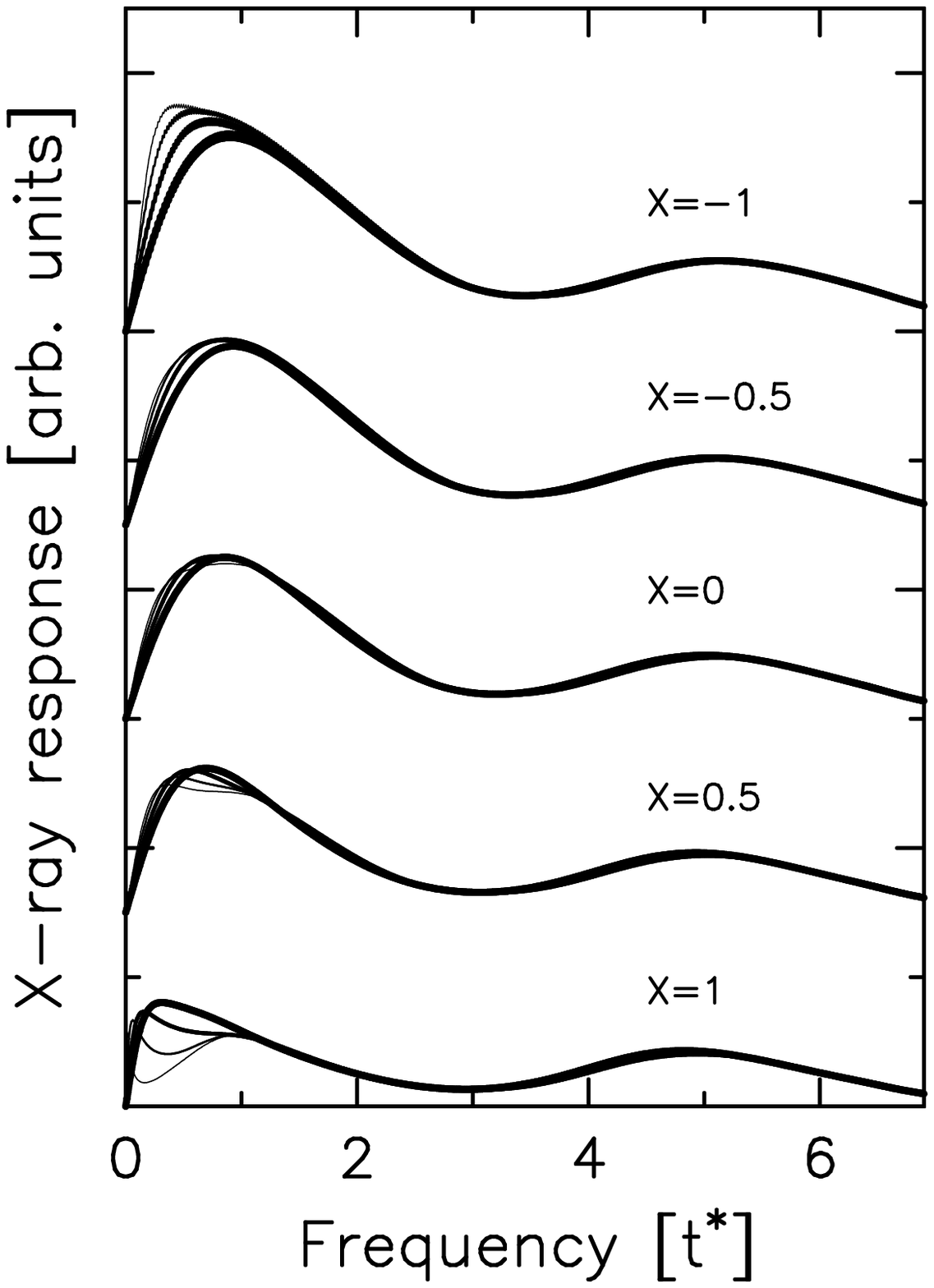}
\caption{
\label{fig: corr_xray}
Inelastic light scattering at $U=4.24$ and $\rho_e=0.8$ for different 
temperatures.  Five values of the transferred photon momentum
are plotted, each shifted by an appropriate amount, and running from the
zone center to the zone boundary along the zone diagonal.  The results
are for the $B_{\rm 1g}$ sector only. The temperature decreases with decreasing
thickness and ranges from 0.127 to 0.077 to 0.047 to 0.029.
}
\end{figure}

We end our results by showing the inelastic light scattering response
of X-rays with momenta that run along the Brillouin zone boundary.
Fig.~\ref{fig: metal_xray} shows the moderately correlated metal and
Fig.~\ref{fig: corr_xray} shows the more strongly correlated metal.
Let's first start with the moderately correlated results in 
panel~\ref{fig: metal_xray}~(a).
There one sees near the Brillouin zone center that the fermi-liquid peak evolves
away from the charge-transfer peak as $T$ is lowered.  But once the photon has
finite momentum, the fermi-liquid feature is strongly suppressed due to
the availability of phase-space to create particle-hole pairs.  We see
no peak separate out, but the scattering response does appear to sharpen
as $T$ is reduced (although it is generically much broader than what is
seen at the zone center).  In panel (b), which is even more weakly correlated,
we find similar behavior except at the zone boundary $(X=-1)$, where the results
are counter-intuitive.  At higher $T$, we see the low-energy X-ray scattering
increase as $T$ decreases, although there is only weak $T$ dependence.
But at the lowest temperature, the response is depleted, as if the system
is becoming more insulating.  This behavior does not seem to occur at 
other momentum transfers, or in the more strongly correlated case, and we
do not have a good understanding for why it happens here, but the scattering
of large momentum photons with correlated charge excitations is a complex
process that is not easy to infer quantitative results about.

Moving on to the more strongly correlated system in Fig.~\ref{fig: corr_xray},
we see some interesting behavior there as well.  Now we can see the remnant 
of the fermi-liquid peak split-off at low energy surviving out to $X=0.5$,
although the peak does not move to lower energy as $T$ is reduced, as it does
at the zone center.  We see the overall broadening of peaks at finite
momentum, and a sharpening of features as $T$ is reduced.  The slope of
the scattering response at the zone boundary can become quite large at low
$T$.

\section{Conclusions}

We have investigated inelastic scattering of light in correlated metals
by solving for the dynamical response functions exactly in the 
infinite-dimensional limit.  We examine the Hubbard model and
thereby are limited to analyzing nonresonant
scattering in the $B_{\rm 1g}$ sector and
along the Brillouin-zone diagonal. The Raman scattering case corresponds to
sitting at the zone center, while the inelastic X-ray scattering allows
one to probe momentum transfers along the zone diagonal.  Since the $B_{\rm 1g}$
response is related to the optical conductivity, many of our results are similar
to those found when investigating the optical conductivity, such as the
appearance of three generic features corresponding to a fermi-liquid
peak, a ``mid-IR'' feature, and a charge-transfer peak.  The relative
weights and temperature dependences of these three features vary with
coupling strength and electron filling, but at low enough temperature, we
will always see a fermi-liquid peak split off and evolve towards zero frequency
as $T$ is lowered (except at half filling with sufficiently large
$U$ that creates an insulator).  When 
we allow the photon to exchange momentum with the
charge excitations, many of the sharp features in the Raman scattering
are blurred, and the peaks are broadened.  We no longer see the fermi peak
evolve at finite momentum transfer, but as the correlations are made
stronger, we do sometimes see remnants of the fermi-liquid peak evolution
(when ${\bf q}$ is close to the zone center).
Near the zone boundary, however, there is just a broad, relatively featureless
response, whose low-energy slope increases as the system is more correlated
and as the temperature is lowered.

To date, there has been limited examinations of electronic Raman scattering or
inelastic X-ray scattering of correlated metallic systems.  The most
studied case is that of the high-temperature superconductors, but they cannot
be examined over the full temperature range because they have intervening
magnetically ordered, or superconducting phases, which can dramatically
change the Raman scattering.  Similarly, most resonant inelastic X-ray
scattering experiments have focused on correlated insulators, because of the 
difficulty in removing the elastic peak from the response function at the
lowest energies, where one might expect interesting fermi-liquid features to 
appear.  Our theoretical results make a number of predictions for how the
temperature, doping, and correlation strength dependence of inelastic light 
scattering varies in correlated metals and we believe it might be possible 
to see some of these features in heavy-fermion materials, which have not yet
been exhaustively studied with these techniques.

\acknowledgments

J.K.F. acknowledges support of the National Science Foundation under grants
DMR-9973225 and DMR-0210717.  T.P.D. acknowledges support from 
the National Research and
Engineering Council of Canada and PREA.  R.B. and Th.P. acknowledge support by 
the Deutsche Forschungsgemeinschaft, through the Sonderforschungsbereich 484.
We also acknowledge useful discussions with S.L. Cooper, R. Hackl,
J.P. Hill, M.V. Klein, and S. Shastry.


\begin{thebibliography}{99}
\bibitem{footnote}
There is a special problem in calculating the inelastic X-ray scattering at
the zone diagonal point corresponding to $X=-1$.  This occurs because there
is a pole in the integrand that lies arbitrarily close to the real
axis.  Hence caution must be used in numerically evaluating the integrand.
The effect of this pole is not too dramatic on the analytic continuation
aside from introducing some extraneous high-frequency oscillations in the
imaginary part of the scattering response function.  The oscillations
are typically small in amplitude and can be corrected by properly handling the
pole term in the numerical integrations.
\end{thebibliography}
\end{document}